# The Impact of Natural External Forcing on Ocean Heat Uptake Efficiency Since the 1980s

Jia-Rui Shi[1], Laure Zanna[1,2], and Alistair Adcroft[3]

[1]*Courant Institute of Mathematical Sciences, New York University, New York, NY, USA*
[2]*Center for Data Science, New York University, New York, NY, USA*
[3]*Princeton University Atmospheric and Oceanic Sciences Program, Princeton, USA*

\* Corresponding author: Jia-Rui Shi ([jia-rui.shi@nyu.edu)](jia-rui.shi@nyu.edu))

**Key Points:**

- The Ocean Heat Uptake Efficiency (OHUE) has notably increased over recent decades, as evidenced by observations and model simulations.
- This increase of OHUE is closely associated with the eruption of Mount Pinatubo (1991).
- Following the eruption, a gradual ocean recovery leads to a peak in OHUE around year 2000.

**Abstract**

We investigate the temporal evolution of ocean heat uptake efficiency (OHUE) using observations and large ensemble model simulations. OHUE, defined as the ratio of the rate in ocean heat uptake to changes in global mean surface temperature anomalies, has exhibited significant variability over recent decades. We found a relatively low OHUE in the late 1980s, a peak around 2000, and a subsequent decline. A key finding is the significant influence of natural external forcing, mainly volcanic eruptions, which causes an abrupt decline in OHUE followed by a gradual recovery. The 1991 Mount Pinatubo eruption, a major volcanic event of the 20th century, had a lasting impact on OHUE. This study emphasizes the contribution of mid-latitudes to global OHUE changes. Our findings underscore the importance of considering natural external forcing in understanding climate dynamics and suggest conducting idealized experiments to quantify the potential effects of future volcanic eruptions on OHUE.

**Plain Language Summary:** We examine how efficiently the ocean has been absorbing heat from the Earth's surface to slowdown the global warming over the past few decades. We found that the efficiency has varied over time, showing a notable increase following the 1991 Mount Pinatubo volcanic eruption, peaking around the year 2000, and then declining. Our analysis of observations and climate model simulations indicates that natural events like volcanic eruptions significantly influence how much heat the oceans absorb. We identified that the mid-latitude regions play a crucial role in driving these variations. Additionally, the Southern Ocean has become less efficient at absorbing heat into the ocean interior ocean in recent periods. These findings emphasize the importance of considering natural external forcing in climate studies to improve the accuracy of future climate predictions.

**1 Introduction**

The ocean plays a central role in regulating the Earth's climate by absorbing more than 90% of the excess energy caused by anthropogenic greenhouse gas emissions (Cheng et al., 2017; Rhein et al., 2013). The extent to which these processes efficiently sequester heat from the ocean surface partially determines the rate of surface warming. Ocean Heat Uptake Efficiency (OHUE, W m$^{-2}$ K$^{-1}$), defined as the rate of ocean heat uptake per degree of change in the global-mean surface temperature (Gregory & Mitchell, 1997; Raper et al., 2002), is critical in moderating the pace of global warming. Therefore, OHUE is a key metric for understanding the ocean's buffering capacity against anthropogenic forcing.

In previous studies, OHUE has been considered as a single number by a given model, quantified in strongly forced scenarios. The factors found to influence or at least correlate with OHUE have been wide-ranging, and include ocean stratification, pycnocline depth, Atlantic meridional overturning circulation, ocean ventilation, mesoscale eddy diffusivity, and vertical diffusivity (e.g., Bourgeois et al., 2022; Kuhlbrodt & Gregory, 2012; Marshall & Zanna, 2014; Newsom et al., 2023; Saenko et al., 2018; Winton et al., 2014). For instance, there are notable correlations between the Atlantic meridional overturning circulation strength and OHUE (Kostov et al., 2014; Winton et al., 2014). This correlation is found to be due to the shared dependence on the transient ocean eddies (Saenko et al., 2018). From a more generalized perspective on regional influences, it has been found that the OHUE is primarily controlled by mid-latitude ventilation strength in the background climate, which is itself related to pycnocline depth and stratification (Newsom et al., 2023). The pycnocline can serve as a proxy for mid-latitude ventilation depth in the background state, and these same ventilation processes contribute most significantly to the global OHUE under anthropogenic forcing.

However, OHUE is not constant over time; it changes in response to a combination of internal ocean processes, external climate forcing, and feedback mechanisms (Gregory et al., 2015). These changes in OHUE may have profound implications for the pace of climate change and the predictability of future warming. Recent studies have highlighted significant changes in OHUE over the historical periods (Cael, 2022; Sohail et al., 2023). For instance, OHUE increased over the past five decades since 1970 by $0.19 \pm 0.04$ Wm$^{-2}$ K$^{-1}$ and was on average $0.58 \pm 0.08$ Wm$^{-2}$ K$^{-1}$ during this period (Cael, 2022). This rate of change has been diagnosed from the observed ocean heat content and global temperature records. Based on climate models, Sohail et al. (2023) focused on the anthropogenic effects over the historical period, such as greenhouse gas and anthropogenic aerosol emissions, on the changes in OHUE. In aerosol-only simulations, the OHUE has declined by $43 \pm 14\%$ since 1980, and is suggested to be linked to the slowdown of aerosol forcing during this period. The associated pattern is characterized by the reduction of OHUE in the tropics and subtropics, while it remains efficient in polar and sub-polar regions (Sohail et al. 2023).

This study focuses on the temporal evolution and external drivers of OHUE change during recent decades in observations and models. By analyzing simulations from the CMIP6 archive and associated single-forcing experiments from DAMIP (Detection and Attribution Model Intercomparison Project), we seek to unravel the mechanisms behind the recent evolution of OHUE and their broader climate implications. The findings aim to advance our understanding of the ocean's role in moderating global warming and contribute to reducing uncertainties in climate projections.

## 2 Materials and Methods

2.1 Observational Datasets

We use two gridded datasets of observed ocean potential temperature. These are datasets from Institute of Atmospheric Physics (IAP) and the subsurface temperature analysis led by Ishii (hereafter Ishii). The IAP ocean potential temperature analysis has a horizontal resolution of 1° × 1° with 41 vertical levels from the surface down to 2,000 m and spans the period 1940 to 2022 (Cheng et al., 2018). The subsurface temperature analysis from Ishii has a horizontal resolution of 1° × 1° with 28 vertical levels from the surface down to 3,000 m and spans the period from 1955 to the present (Ishii et al., 2005). The upper 2,000 m OHC is calculated based on these datasets.

Additionally, we use a reconstruction based on the Green's Function (GF) method which estimates the ocean temperature field (Zanna et al., 2019; Wu et al., accepted). The GF method utilizes a kernel derived from ocean tracer observations and surface "excess" temperature as the boundary condition, where "excess" excludes SST changes due to ocean circulation change. There are 12 GF kernels estimated from ocean tracer observations, 3 observational datasets for global SST, and 2 excess temperature fields from simulations, resulting in 72 members for the GF OHC estimate. More details are in Wu et al. (2022) and Wu et al. (accepted).

For surface air temperature, we use the HadCRUT5 dataset, a widely used global temperature record that combines both in situ and satellite-based measurements (Morice et al., 2021). Covering the period from 1850 to the present, HadCRUT5 allows for a thorough comparison of model-derived surface temperature changes over the historical period.

2.2 Large-Ensemble Simulations and Single-Forcing Experiments

This study uses the outputs from Phase 6 of the Coupled Model Intercomparison Project (CMIP6), focusing on simulations from 6 selected models (ACCESS-ESM1-5, CNRM-CM6-1, CanESM5, IPSL-CM6A-LR, MIROC6, MPI-ESM1-2-LR) that have large ensemble simulations (≥ 25 members; Eyring et al., 2016). We also utilize the 50-member sub-ensemble, which follows the original CMIP6 biomass burning protocol, of the CESM2 large ensemble simulations (CESM2LE; Rodgers et al., 2021), in this study. These 7 models together provide outputs from the historical all-forcing (HIST) simulations, which include all major external forcings. In addition, the study also examines the results from single-forcing experiments from DAMIP (Gillett et al., 2016) from these models, to which isolate the impact of individual forcings, namely greenhouse gases (GHG), anthropogenic aerosols (AER), and natural forcings like volcanic aerosols and solar irradiance (NAT) or everything else evolving in CESM2LE (Simpson et al., 2023). The number of realizations from each model is provided in Supporting Information (Table S1). These experiments aim to quantify the contributions of each forcing to changes in OHUE.

We utilize the ocean potential temperature and surface air temperature at 2 m from models. The ocean heat content (OHC) in the upper 2,000m is obtained by vertically integrating the temperature field and multiplying by reference values of density (1,025 kgm$^{-3}$) and specific heat (4,000 J kg$^{-1}$ K$^{-1}$). Model drift in OHC is removed by subtracting a cubic polynomial fit based on the preindustrial control run over its full length in time for each model (Irving et al., 2020). All model output has been interpolated to a regular 1° × 1° grid.

2.3 Ocean Heat Uptake Efficiency

Ocean Heat Uptake Efficiency (OHUE) is defined as:

$$OHUE = \frac{\Delta N}{\Delta T} \qquad (1)$$

where $\Delta N$ is the global-mean rate of ocean heat uptake, and $\Delta T$ is the global-mean surface air temperature change (GMST). The unit of OHUE is in W m$^{-2}$ K$^{-1}$, representing the efficiency with which the ocean absorbs heat relative to changes in GMST. We use the long-term trend of upper 2,000 m ocean heat content as introduced above to estimate $\Delta N$. This approach is based on the assumption that nearly all the heat storage change occurs within the upper 2,000 m which allows for a comparison between observations and models. The results presented in the main text are based on a 20-year window for calculating the OHC tendency. We also conduct a sensitivity test varying the length of the window. The OHC and GMST are bulk global metrics. We further apply this approach to access the contributions from different external forcings and regions (see Sections 3.2, 3.3).

**3 Results**

**3.1 Historical Evolution of Ocean Heat Uptake Efficiency**

The historical ocean heat uptake efficiencies (OHUE), obtained over two-decade time windows from observations and CMIP6 historical runs, are shown in Figure 1a. The OHUE from observations is within the range of 0.4–1.0 W m$^{-2}$ K$^{-1}$. The spread of results based on the Green's Function (GF) method is approximately 0.3–1.3 W m$^{-2}$ K$^{-1}$, influenced by the choice of SST patterns and the prior estimate of the GF. The simulated OHUE from 7 models with large-ensemble simulations is found to be approximately 0.3-1.0 W m$^{-2}$ K$^{-1}$, and matches the observed OHUE magnitude. The historical OHUE in this study is consistent with previous estimates based on observations and models (Cael, 2022; Kuhlbrodt & Gregory, 2012; Newsom et al., 2023; Sohail et al., 2023).

Importantly, there is a remarkable variation in OHUE over recent decades (Figure 1a): the value is relatively low during the late 1980s, rose to peak around the year 2000 at 0.8 W m$^{-2}$ K$^{-1}$ from GF, and then decreased back to around 0.6 W m$^{-2}$ K$^{-1}$ by 2011 (i.e. 2001-2020 period). It indicates that the ocean has just passed its most efficient period for absorbing heat from the atmosphere per unit of global surface temperature increase. The ensemble mean results from the models can capture this temporal feature. We also conducted a sensitivity test regarding the window length used to calculate OHUE (see Figure S1 in Supporting Information). The results from 15-year and 25-year windows are consistent with those shown in Figure 1, and longer windows naturally yield smoother results.

We show the two results centered at the years 1988 and 2000 to evaluate whether the changes in OHUE are statistically significant between these periods. We found that 6 out of 7 models show a significant increase, with p-values less than 0.01, except for model CNRM-CM6-1 (Figure 1b). Similarly, the observed OHUE change between the 1988 and 2000 periods based on GF is also significant. The increase in OHUE is $0.29 \pm 0.10$ Wm$^{-2}$ K$^{-1}$, based on the multi-model mean of six models and their standard deviation. Moreover, the relative change and the corresponding cross-model uncertainty is +67% ± 35%. These results from the multi-model ensemble mean illustrate that the temporal feature reflects a forced response. The observed OHUE increase is 0.13 (+25%), 0.11 (+20%), and 0.23 (+41%) Wm$^{-2}$ K$^{-1}$ from the IAP dataset, Ishii dataset, and GF, respectively. In addition, the decrease between 2010 and 2000 based on GF is also significant (not shown).

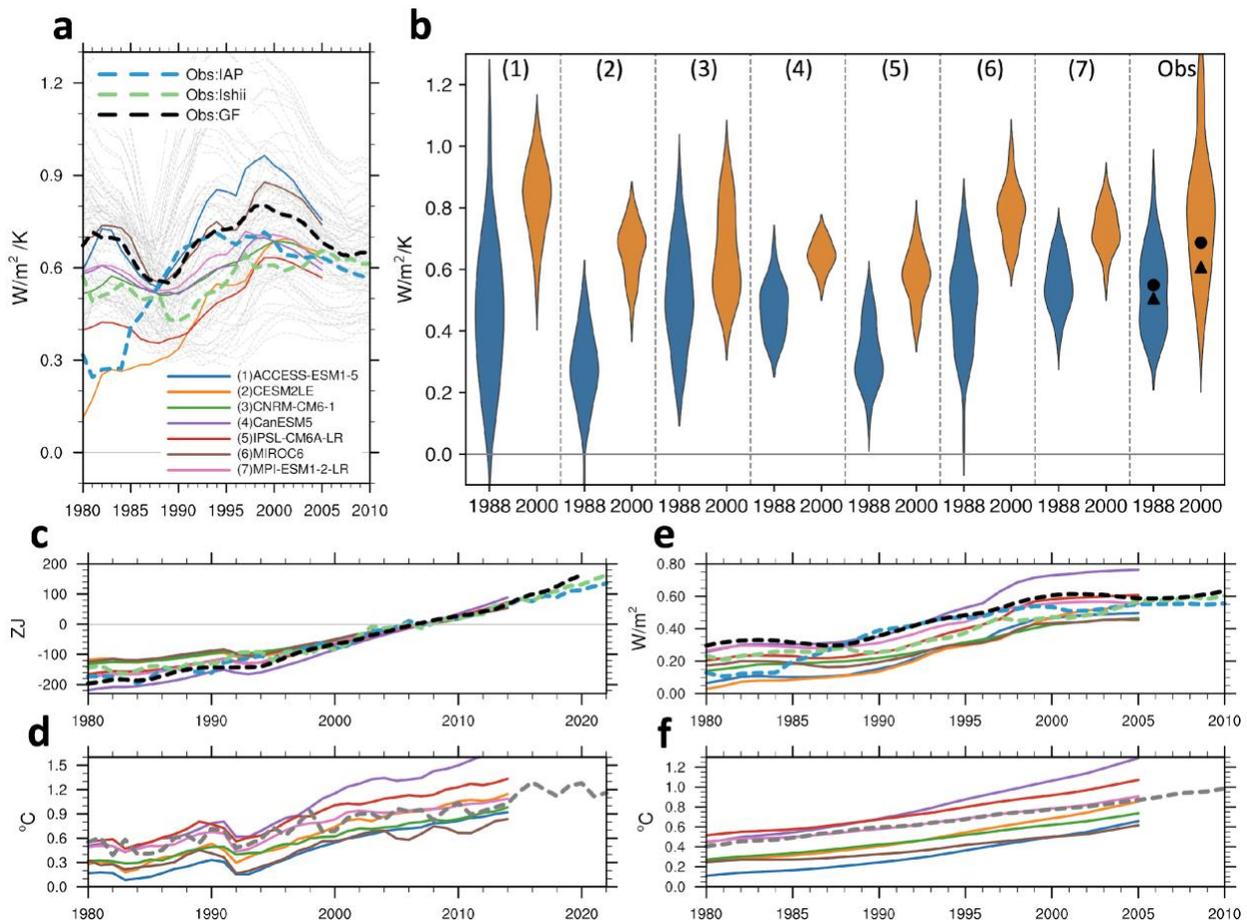

**Figure 1**. Time evolution of ocean heat uptake efficiency (OHUE) and associated variables from observations and CMIP6 large-ensemble simulations. (a) Time evolution of OHUE over a 20-year sliding window. The x-axis shows the centered year of the 20-year window. The results represent the ensemble mean from each model. (b) Distributions of OHUE from 1978-1997 (centered at 1988) and 1990-2009 (centered at 2000). The number in parentheses refers to the CMIP6 model (see Table S1 in Supporting Information). For the observations (Obs), the distribution is from 72 members from GF, circles represent IAP data, and triangles represent Ishii data. (c,d) Time evolution of global ocean heat content (OHC) and global mean surface temperature (GMST), respectively. The dashed gray curve is the GMST from HadCRUT5. (e) Ocean heat uptake rate, calculated as the 20-year trend of OHC. (f) 20-year running mean of GMST. As is panel (a), the x-axis in (e,f) indicates the centered year.

The ocean heat content (OHC) and global mean surface temperature (GMST) anomalies have steadily increased since 1980, with temporary dips caused by the volcanic eruptions of El Chichón (1982) and Pinatubo (1991) (Figures 1c,d). The ocean heat uptake rate ($\partial OHC/\partial t$; Figure 1e), compared with the nearly linear increase in the 20-year running mean of the GMST anomaly ($\Delta T$; Figure 1f), plays a dominant role in driving the recent variation in OHUE. Therefore, we next focus on the variations in the ocean heat uptake rate in response to different external forcings, such as volcanic eruptions, greenhouse gas (GHG), and anthropogenic aerosol (AER) emissions.

### 3.2 Contributions from Individual Forcings

The historical OHC can be represented as a linear combination of responses to GHG, AER, and natural forcings (Sohail et al., 2023). While the mean surface air temperature is influenced by various forcings, our focus here is solely on partitioning the ocean heat uptake rate ($\Delta N$) into components driven by these forcings to quantify their respective contributions to the overall OHUE change. In Eq. 2, the subscripts HIST, GHG, AER, and NAT refer to results from all-forcing, greenhouse gas-only, aerosol-only, and natural-external-only forcing experiments, respectively. The denominator is the $\Delta T$ anomaly from the historical all-forcing run (Figure 1f).

$$\frac{\Delta N_{\text{HIST}}}{\Delta T_{\text{HIST}}} = \frac{\Delta N_{\text{GHG}} + \Delta N_{\text{AER}} + \Delta N_{\text{NAT}} + \text{Res}}{\Delta T_{\text{HIST}}} \qquad (2)$$

The main goal of this section is to identify which external forcing leads to the temporal features in historical OHUE. Although the GHG-induced global OHC continues to increase, its ratio to the $\Delta T$ (i.e., OHUE) is gradually decreasing over time (Figure 2a). The aerosols give rise to negative OHUE associated with its cooling effect in the ocean (Boland et al., 2022; Shi et al., 2023), which offsets some of the ocean warming from GHGs (Figure 2b). Both the GHG- and AER-induced changes are quite linear, without exhibiting the observed variation. The NAT-only experiment shows a distinct drop between 1985 and 1990, followed by an increase thereafter (Figure 2c), closely matching the observed temporal feature. The linear combination of these components imprints the natural forcing effects resulting from the compensation between GHG- and aerosol-driven changes. The decrease after the year 2000 is stronger in the combined OHUE than in the NAT experiment (Figure 2c,d).

Volcanic aerosol loading is the dominant driver of natural external forcing, with the 1991 Mount Pinatubo eruption standing out as a particularly significant event. This eruption, much stronger than other recent volcanic events, had notable impacts on the Earth's energy imbalance and ocean heat uptake (Gleckler et al., 2006; Allan et al., 2014; Santer et al., 2019; Smith et al., 2016). The OHC anomaly and vertical temperature profile (Figures 2e,f) reveal a sharp decline in temperature following the eruption and the penetration of cooling signal to the deeper layer, reaching its minimum around 1993, followed by a slower recovery phase (Santer et al., 2007; Takahashi & Watanabe, 2016). The relaxation time scale is approximately 10 years for the surface and upper ocean, while it extends from several decades to a century associated with the penetration of cold anomalies to the deeper ocean layers (Stenchikov et al., 2009; Dogar et al., 2020). Results across all models consistently show this temporal evolution of ocean heat uptake to natural external forcing (Figure S2 in Supporting Information). Notably, the Pinatubo-induced ocean heat uptake mainly contributes to the ocean warming rate (Figure 1e) and the recent variation in the OHUE (Figures 1a, 2c,d), reflecting a long-lasting volcanic signal in the ocean and its impact on the climate system.

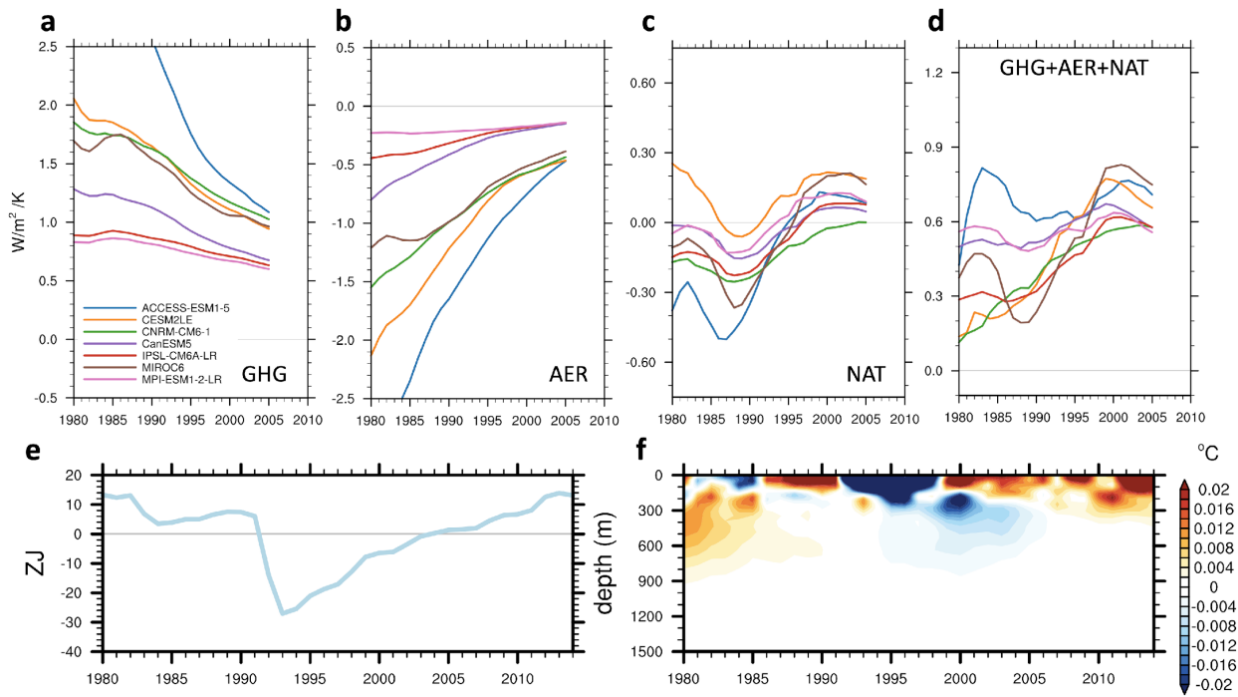

**Figure 2.** Contributions of single forcings to OHUE and ocean temperature during the historical period. (a,b,c) OHUE calculated using the ocean heat uptake rate from single-forcing experiments: greenhouse gases (GHG), aerosols (AER), and natural forcing (NAT). (d) Linear combination of OHUE derived from three single-forcing experiments. (e,f) Global OHC anomaly and vertical profile of global mean ocean temperature anomaly relative to the 1980-2014 mean from MIROC6 model from NAT-only experiment.

### 3.3 Regional Responses

The evolution of OHUE can be decomposed into different latitude bands to find their contributions to global change. For example, the latitudinal components of OHUE can be expressed as:

$$OHUE_{\text{lats}_i} = \frac{1}{A_G} \frac{\int_{y_{i0}}^{y_{i1}} \int \Delta N(x,y)\, dx\, dy}{\Delta T_G} \qquad (3)$$

In this equation, $A_G$ and $\Delta T_G$ represent the global surface area and the GMST anomaly, respectively. $\Delta N(x,y)$ is the regional heat uptake rate. The global sum of $OHUE_{lat}$ equals the global OHUE, as discussed previously. Here, we focus on the regional contributions of ocean heat uptake to the changes in global OHUE.

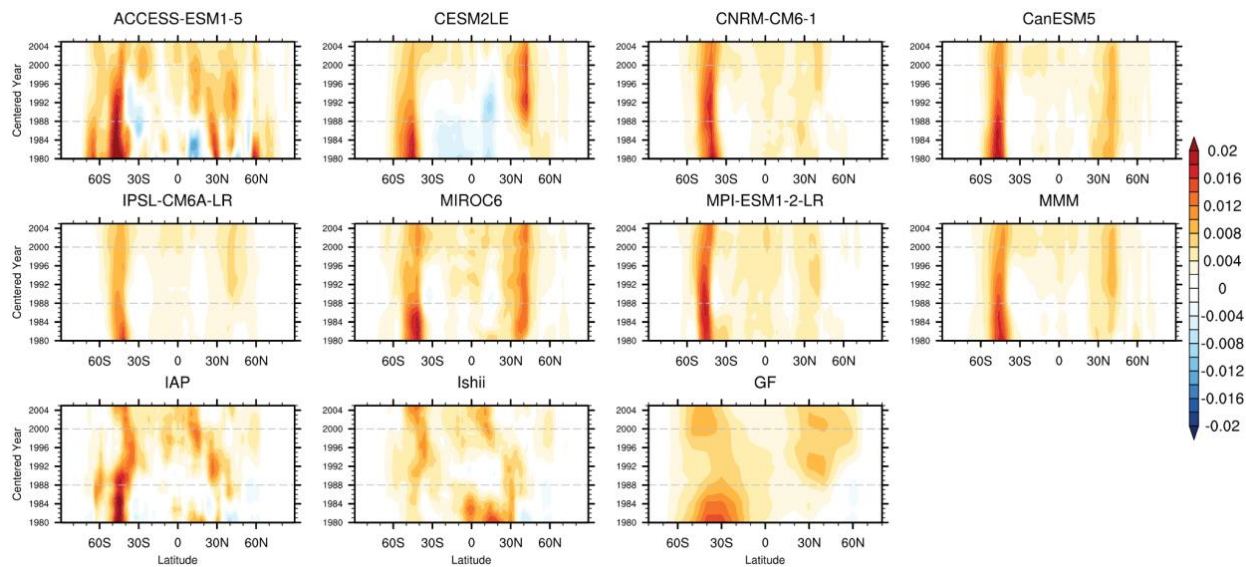

**Figure 3.** OHUE as a function of time and latitude from models and observations. The y-axis shows the centered year of the 20-year window. The unit is Wm$^{-2}$ K$^{-1}$ per 1° latitude band. Top two rows are OHUE from the ensemble-mean of individual models and the multi-model mean (MMM). The bottom row shows the results from IAP, Ishii datasets, and Green's Function (GF) method. The regional OHUE based on GF is obtained by using excess temperature without effect of ocean circulation change.

The mid-latitudes play a leading role in contributing to the time-mean global OHUE, with consistency shown across models and datasets (Figure 3). For instance, two strong bands of positive values are located in the Southern Ocean, particularly between 40˚S and 55˚S, and between 30˚N and 45˚N in the Northern Hemisphere. The subtropical North Atlantic and the region to the north of the Kuroshio Front both contribute to responses in the Northern Hemisphere (Figure S3 in Supporting Information). These results indicate a substantial heat gain in the mid-latitude ocean basins, compared to the rest of the global ocean (Frölicher et al., 2015; Shi et al., 2018; Zanna et al., 2019; Newsom et al., 2020), more efficiently mitigating the global surface warming.

Regarding the long-term change, the decrease of regional OHUE is evident in the Southern Ocean from the ensemble mean of each model (Figure 3). It suggests that the Southern Ocean has become less efficient in taking up heat in recent decades. In the Northern Hemisphere, some models show an increase in regional OHUE (CESM2LE, IPSL-CM6A-LR, and MPI-ESM1-2-LR), while some (such as CanESM5 and MIROC6) show no clear direction of change. The multi-model mean results show a slight increase since 1980, followed by a plateau starting in the 1990s. This behavior may be linked to the anthropogenic aerosol effect on the Atlantic meridional overturning circulation (AMOC; e.g., Shi et al., 2018). In the tropics (i.e. 30˚S-30˚N), there is an upward trend based on models.

OHUE estimated from observational datasets shows both similarities and differences compared to the simulated results, and these observational results even exhibit distinct spatial features. For instance, the increase of regional OHUE between the 2000 period and 1988 period in the tropical regions is evident in both IAP and Ishii datasets (Figures 3 and 4). Changes in the Southern Ocean are more consistent between these two datasets after the 1990s. However, the Ishii data appears to have a much smaller OHUE in the early periods compared to the IAP data, which may be attributed to sparse observations in the Southern Ocean (Swart et al., 2018; Shi et al., 2021). In the Northern Hemisphere mid-latitudes, neither dataset shows a clear long-term trend in OHUE. The results based on GF reflect the excess temperature changes and the associated OHUE, excluding the redistribution of heat due to changes in ocean circulation. Therefore, there is a substantial difference between the two datasets and model simulations when investigating the regional features. The excess OHUE derived using GF shows a notable contribution from the mid-latitudes. The broader distribution in GF, compared with the other datasets and models, is primarily due to its coarse resolution (10° in latitude). The enhanced OHUE in the Northern Hemisphere is more consistent with the MMM results (Figures 3 and 4). This increase mainly occurs in the North Atlantic (Figure S4 in Supporting Information). The dipole feature observed in the IAP and Ishii datasets (Figure S4) suggests that ocean circulation changes, such as the AMOC, also play a role in this ocean basin.

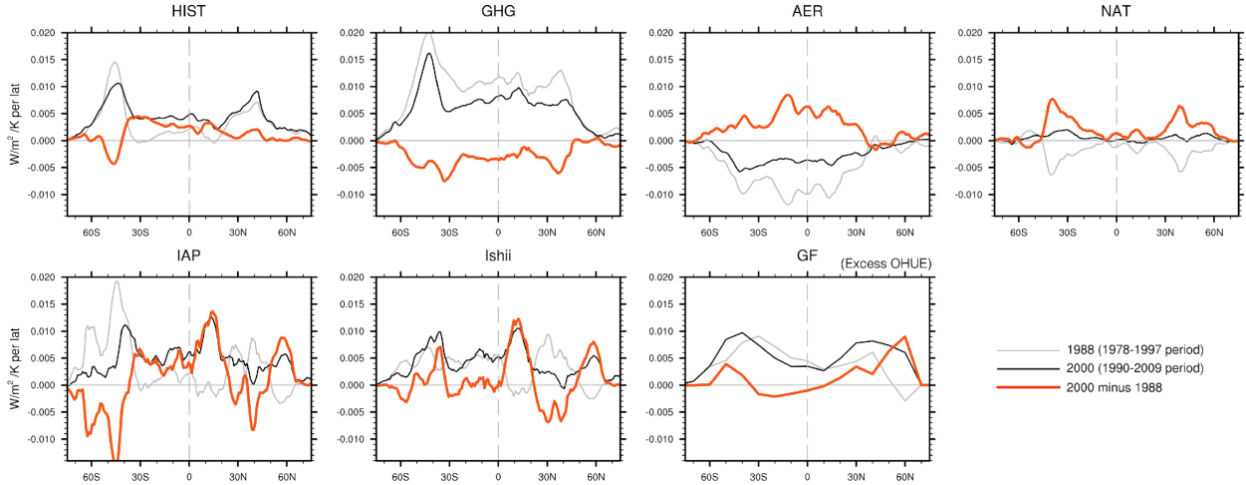

**Figure 4.** OHUE change between the 2000 period and 1988 period. The top row shows the OHUE change from multi-model mean from HIST, GHG, AER, and NAT, respectively. The bottom row shows the results from IAP, Ishii datasets, and as well as the excess OHUE based on GF.

The historical results can be decomposed into contributions from GHG, AER, and NAT forcing (Figure 4). In both GHG and AER, OHUE weakens (i.e., moves closer to zero) from the 1988 period to the 2000 period, as shown in Figures 2a and 2b. The positive response in AER is more pronounced in the tropical latitudes, which could be linked to the geographic distribution of anthropogenic aerosol forcing. Although natural external forcing is often overlooked in long-term climate change due to its smaller magnitude relative to GHG and AER forcings, its effect on regional OHUE change is comparable to that of the other two forcings (Figure 4d). The zonal pattern indicates that the NAT-forced change is mainly affected by the ocean heat uptake in both mid-latitudes, which resembles the GHG-induced pattern (Figure 4b). In addition, the positive change in NAT is primarily due to the ocean temperature decrease during the Pinatubo eruption period (gray curve in Figure 4d).

The OHUE change maps from the IAP and Ishii datasets reveal a strong positive change in the western Pacific and Indian Ocean, along with a negative change in the eastern Pacific—features not present in the ensemble mean of the models (Figure S4 in Supporting Information). This discrepancy between observations and models suggests that internal climate variability in the observations could play a key role in driving these differences. To validate this hypothesis, we examine the results from individual members of the CESM2LE model as an example, which combines the forced response with a random-phase background noise (Figure S5 in Supporting Information). Some realizations match the observed pattern, and some also exhibit a similar pattern but with the opposite sign of change. It is worth noting that internal variability is larger than the forced responses (Figure S6 in Supporting Information), particularly in certain models (ACCESS-ESM1-5 and CNRM-CM6-1). This large internal variability presents a challenge in extracting the forced response from the spatial maps.

**4 Summary and Discussions**

In this study, we investigate the historical evolution of ocean heat uptake efficiency (OHUE), defined as the ratio of ocean heat uptake rate to changes in global mean surface temperature (GMST), using both observations and CMIP6 model simulations. We observed a relatively low OHUE in the late 1980s, a peak around 2000, and a subsequent decline until now. Simulations from seven models with large-ensemble simulations generally matched observational OHUE variation from in-situ measurements and based on a Green's Function approach, confirming an externally forced response in OHUE.

One key finding is the significant influence of natural external forcings, particularly volcanic eruptions, on the temporal variability in OHUE. Specifically, the 1991 Mount Pinatubo eruption caused a significant decline in OHUE, followed by a recovery period, indicating a lasting volcanic effect on ocean heat uptake and its efficiency. The changes in OHUE due to greenhouse gas (GHG) and aerosol (AER) forcings are more linear, largely compensating each other, and do not exhibit the observed change. Notably, the impact of natural external forcings on OHUE has been largely overlooked in previous studies. The long-lasting impacts of volcanic events on OHUE underscore the importance of considering natural variability in understanding climate dynamics.

The substantial increase in global OHUE since the 1980s found in this study confirms previous findings based on different methods (Cael, 2022; Watanabe et al., 2013). In this study, we show that the observed global OHUE increase from 1988 to 2000 by 0.13 (+25%), 0.11 (+20%), and 0.23 (+41%) $Wm^{-2} K^{-1}$ from the IAP dataset, Ishii dataset, and GF, respectively. Watanabe et al. (2013) used the global energy balance equation to estimate observed ocean heat uptake, given the radiative forcing and climate feedback parameters estimated from CMIP5 models. The observed OHUE increased from approximately 0.6 $W m^{-2} K^{-1}$ during 1971–2000 to around 2.1 $W m^{-2} K^{-1}$ during 1991–2012. This increase (+1.5 $W m^{-2} K^{-1}$) appears somewhat overestimated compared to other studies which report a range of 0.5–1.0 $W m^{-2} K^{-1}$ (Kuhlbrodt & Gregory, 2012; Newsom et al., 2023). Cael (2022) integrated the GMST rather than calculating the trend of OHC to estimate OHUE, assuming that the rate of increase remained constant over the integration period. The magnitude of change reported by Cael (2022), +0.83% per year since 1970 with a base value of 0.48 $W m^{-2} K^{-1}$, is similar to our findings. While the increase in OHUE has been strong during certain periods, we show that it has already passed its peak value and is now trending downward. This temporal variation is also observed in Sohail et al. (2023; Figure 2f herein), though their focus is on the effect of anthropogenic aerosols, without considering the role of natural external forcing.

The global OHUE derived from observations falls within the range of model results (Figure 1). The inter-model variability in the ocean warming rate and GMST anomaly appears to be influenced by differences in equilibrium climate sensitivity (ECS). For example, CanESM5, with the largest ECS (5.6°C from Meehl et al., 2020), provides the highest values in warming rate and GMST anomaly by the end of the analysis year, and MIRCO6, with the smallest ECS (2.6°C), shows the lowest values (Figure 1e,f). However, the effect of ECS on OHUE is not as evident (Figure 1a). Although ECS is an important factor in determining the ocean warming rate and GMST anomaly, its influence on OHUE appears to be less pronounced, potentially due to other oceanic processes such as circulation or heat distribution that vary across models. Further analysis is needed to understand the underlying mechanisms.

The vertical profile of ocean temperature responses (Figure S2) also shows some inter-model differences. For instance, the penetration of the volcano-eruption-related cooling signal is shallower in CanESM5 compared to the other models (Figure S2 in Supporting Information). This leads to a smaller magnitude of OHUE variation in CanESM5 (Figure 1), which may be related to differences in vertical diffusivity. Further investigation is needed to better understand the role of mixing processes in shaping OHUE responses, and in general more theory is needed to understand the timescale of response of OHUE to forcing (akin., Marshall and Zanna, 2014).

Regarding the spatial pattern of OHUE evolution, the IAP and Ishii show notable differences before the 1990's (Figure 3), which are linked with the sparseness of earlier records (Swart et al., 2018). The models show a response more similar to IAP, but this does not necessarily imply that it is accurate. We would need more accurate and long-term observational data, particularly in under sampled regions like the Southern Ocean, to provide more robust validation and improve model projections. In addition, given the large internal variability and the relatively small signal-to-noise ratio during the historical period, it would be beneficial to investigate model differences using strong forcing scenarios and a much larger ensemble of models in future studies.

As shown in both the data and models, OHUE has already exhibited a noticeable decrease in the early 21st century (Figure 1a). It is mainly due to reduced efficiency in the Southern Ocean (Figure 3). When considering the moderate emissions scenario (Shared Socioeconomic Pathways 2-4.5), almost all models (except CESM2LE) show a linear decrease in OHUE through to the end of the 21st century (Figure S7 in Supporting Information). This decline is potentially influenced by the deep ocean warming up and increased stratification, which hinders the penetration of excess heat (Gregory et al., 2015; Li et al., 2020; Bourgeois et al., 2022). To further understand the impact of future volcanic eruptions on OHUE and the climate system, idealized experiments simulating large volcanic events combined with future emission scenarios should be conducted. These experiments could help quantify the potential long-lasting effects on ocean heat uptake and provide insights into the interplay between different forcing mechanisms.

**Acknowledgments**

This project is supported by Schmidt Sciences, LLC. This research is also supported in part through the NYU and WHOI High Performance Computing resources. We acknowledge the World Climate Research Programme's Working Group on Coupled Modelling, which led the design of CMIP6 and coordinated the work, and we also thank individual climate modeling groups (listed in Table S1) for their efforts in performing all model simulations analyzed here. We also would like to thank Lijing Cheng, Masayoshi Ishii, Jonathan Gregory, and Quran Wu for their efforts in producing and maintaining the observational datasets utilized in this work.

**Data Availability Statement**

Outputs from CMIP6 and DAMIP experiments are available on the Program for Climate Model Diagnostics and Intercomparison's Earth System Grid (https://esgf-node.llnl.gov/search/cmip6/). The outputs from the Community Earth System Model 2 Large Ensemble Community Project and its Single Forcing project are available at: https://www.cesm.ucar.edu/community-projects/lens2 and https://www.cesm.ucar.edu/working-groups/climate/simulations/cesm2-single-forcing-le, respectively. IAP data are available at: https://climatedataguide.ucar.edu/climate-data/ocean-temperature-analysis-and-heat-content-estimate-institute-atmospheric-physics. Ishii data are available at: https://www.metoffice.gov.uk/hadobs/en4/. HadCRUT5 data are available at: https://www.metoffice.gov.uk/hadobs/hadcrut5/. Ocean heat uptake rates estimated from Green's Function are available at: https://doi.org/10.5281/zenodo.11107298.

Supporting Information for

# The Impact of Natural External Forcing on Ocean Heat Uptake Efficiency Since the 1980s

Jia-Rui Shi[1*], Laure Zanna[1,2], Alistair Adcroft[3]

[1]Courant Institute of Mathematical Sciences, New York University, New York, NY, USA
[2]Center for Data Science, New York University, New York, NY, USA
[3]Princeton University Atmospheric and Oceanic Sciences Program, Princeton, USA

*Corresponding author: Jia-Rui Shi

Email: jia-rui.shi@nyu.edu

**This PDF file includes:**

    Tables S1
    Figures S1 to S7

**Table S1.** CMIP6 models and the number of realizations used in this study. The left column shows the names of 7 models. The models are numbered to match Figure 1b. The rest columns show the number of realizations in the historical all-forcing and single-forcing experiments. For CESM2LE model, the "everything else evolving" experiment is classified as NAT here.

| Model Names | HIST | GHG | AER | NAT |
|---|---|---|---|---|
| (1) ACCESS-ESM1-5 | 30 | 3 | 3 | 3 |
| (2) CESM2LE | 50 | 15 | 15 | 15 |
| (3) CNRM-CM6-1 | 30 | 3 | 3 | 3 |
| (4) CanESM5 | 25 | 15 | 15 | 15 |
| (5) IPSL-CM6A-LR | 32 | 10 | 10 | 10 |
| (6) MIROC6 | 50 | 10 | 10 | 10 |
| (7) MPI-ESM1-2-LR | 50 | 20 | 20 | 20 |

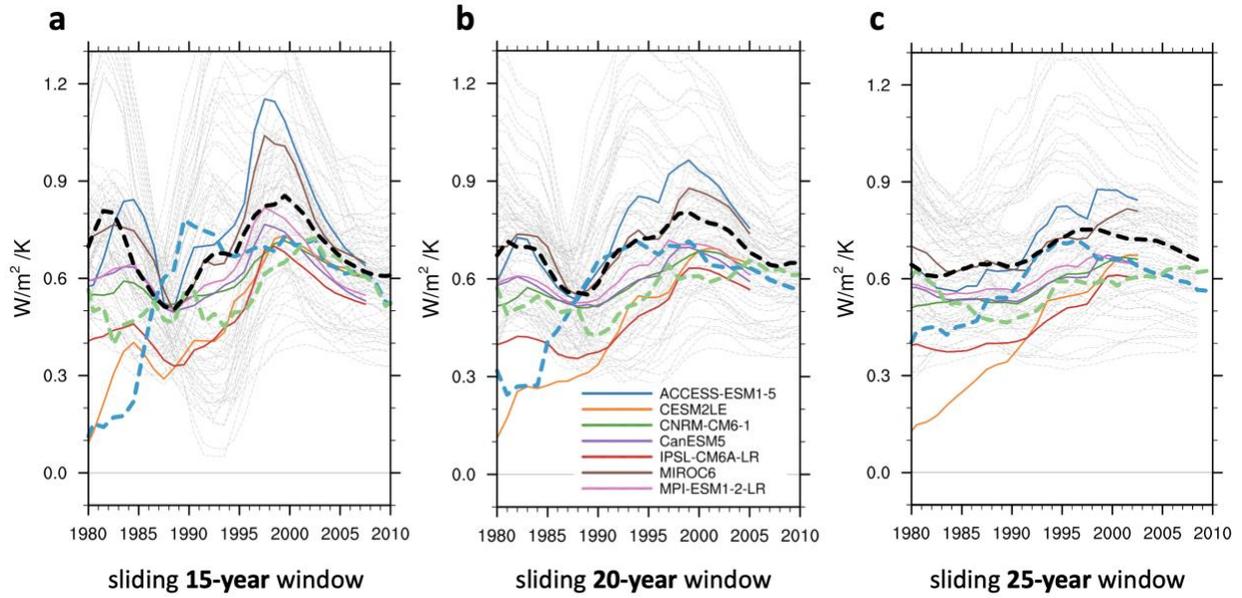

**Figure S1.** Time evolution of ocean heat uptake efficiency (OHUE). (a-c) show observed results over 15-year, 20-year, and 25-year sliding window. Similar to Figure 1a, the solid curves are from 7 model ensemble mean, and dashed ones are from observations.

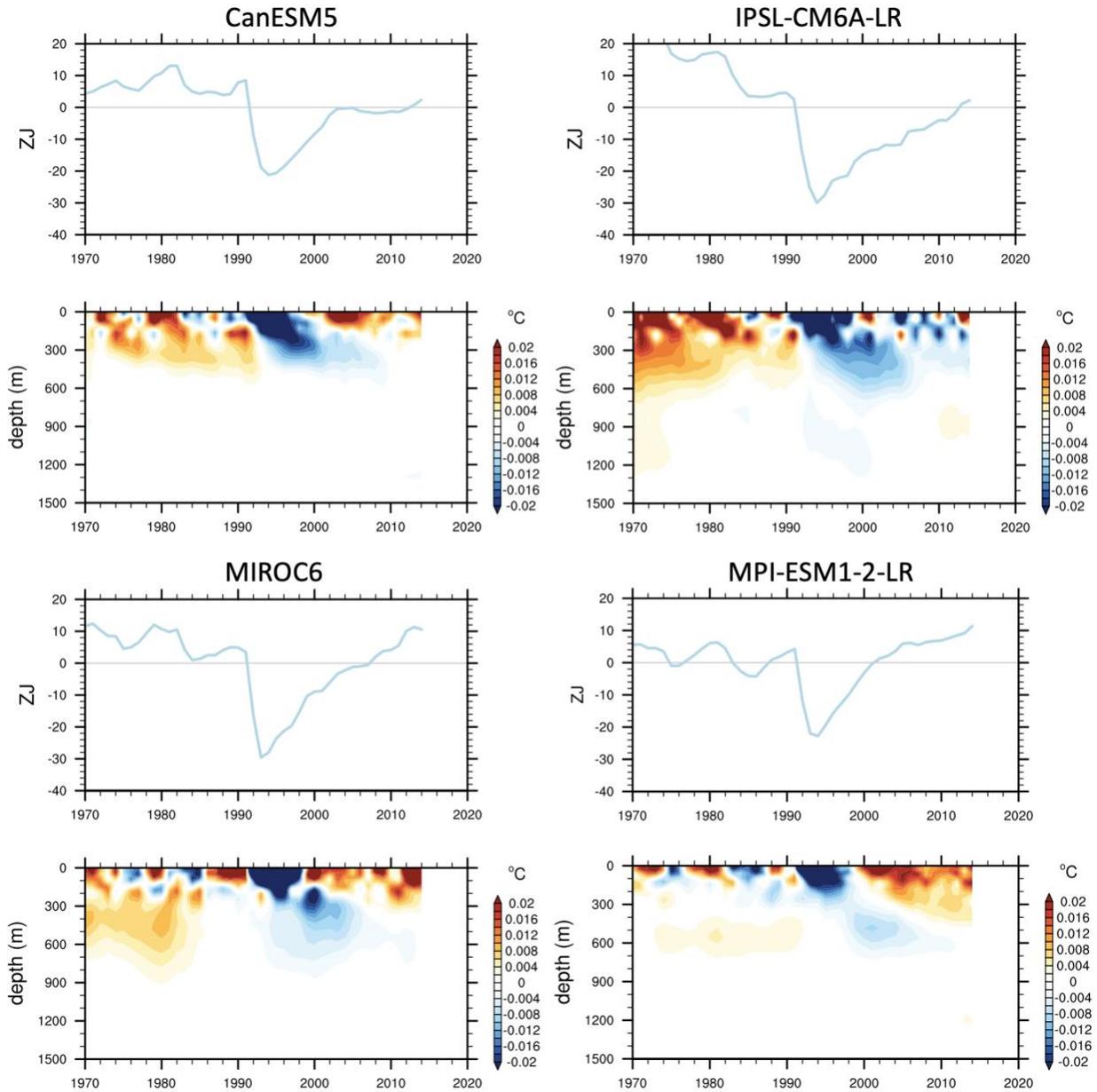

**Figure S2.** Global OHC anomaly and vertical profile of global mean ocean temperature anomaly relative the 1980-2014 mean from NAT experiment.

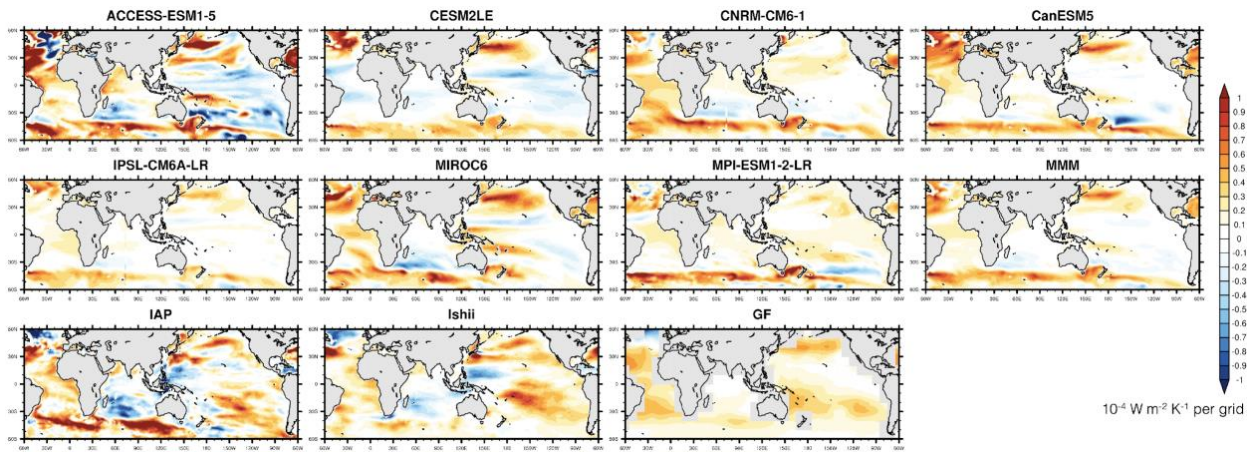

**Figure S3.** Spatial pattern of OHUE averaged within 1978-1997 period (centered at 1988) from models and observations. Top two rows are OHUE from ensemble mean of individual models and the multi-model mean. The unit is $10^{-4}$ W m$^{-2}$ K$^{-1}$ per 1°x1° latitude and longitude.

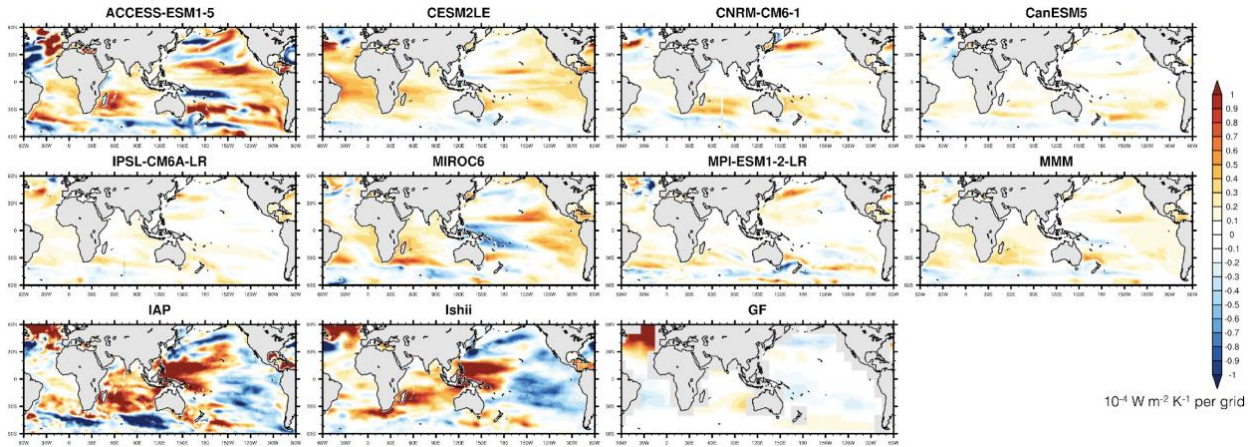

**Figure S4.** Spatial pattern of OHUE change between 2000 period and 1988 period. The unit is $10^{-4}$ W m$^{-2}$ K$^{-1}$ per 1°x1° latitude and longitude.

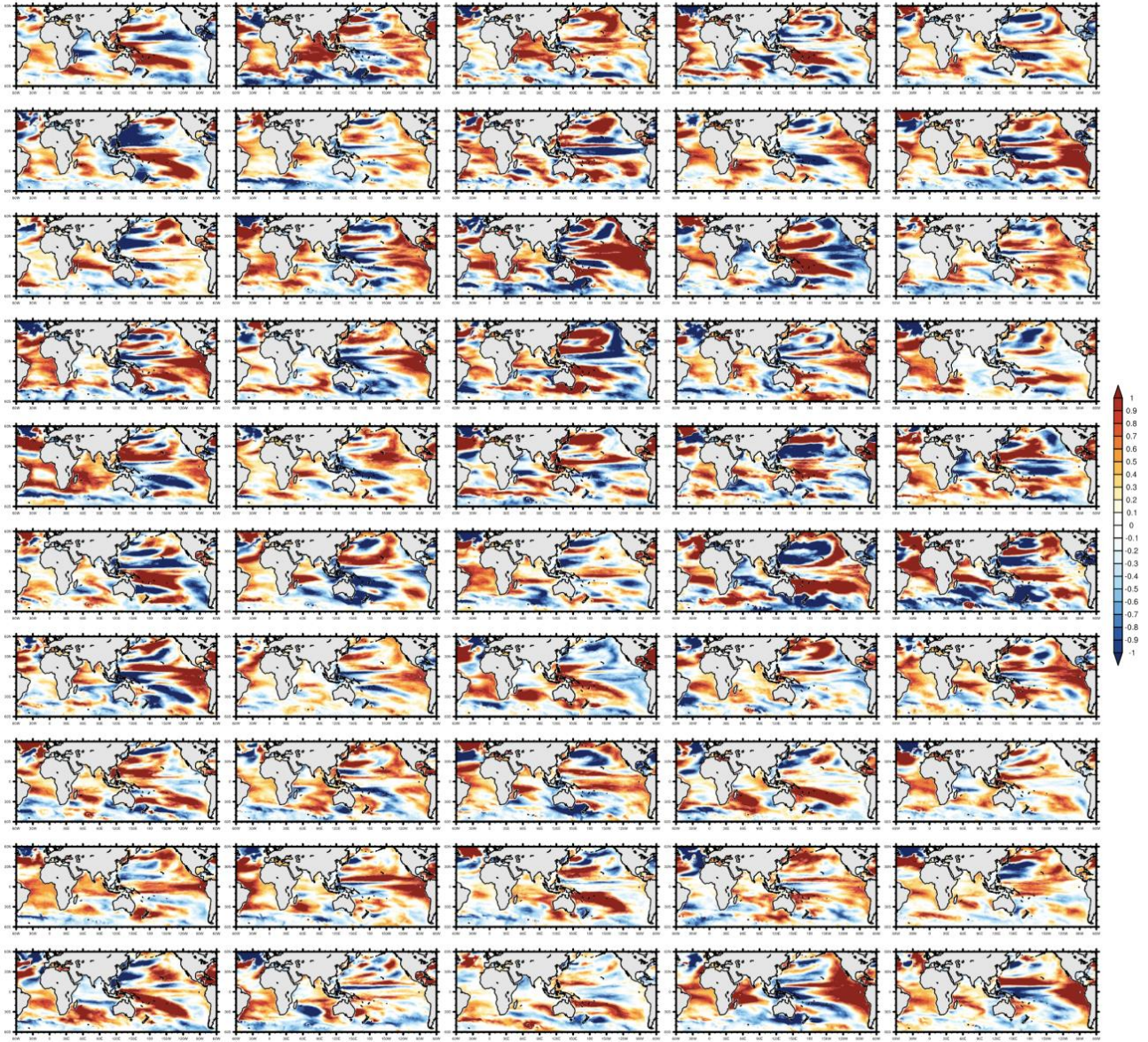

**Figure S5.** Similar to Figure S4, but results are from OHUE change from each member of CESM2LE model. The unit is $10^{-4}$ W m$^{-2}$ K$^{-1}$ per 1°x1° latitude and longitude.

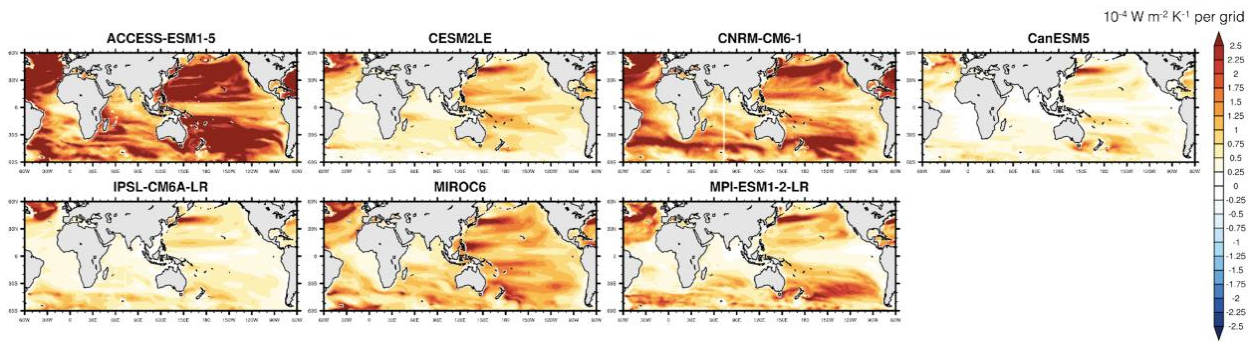

**Figure S6.** Standard deviation of OHUE change between 2000 period and 1988 period across 25 members of each model. The unit is $10^{-4}$ W m$^{-2}$ K$^{-1}$ per 1°x1° latitude and longitude.

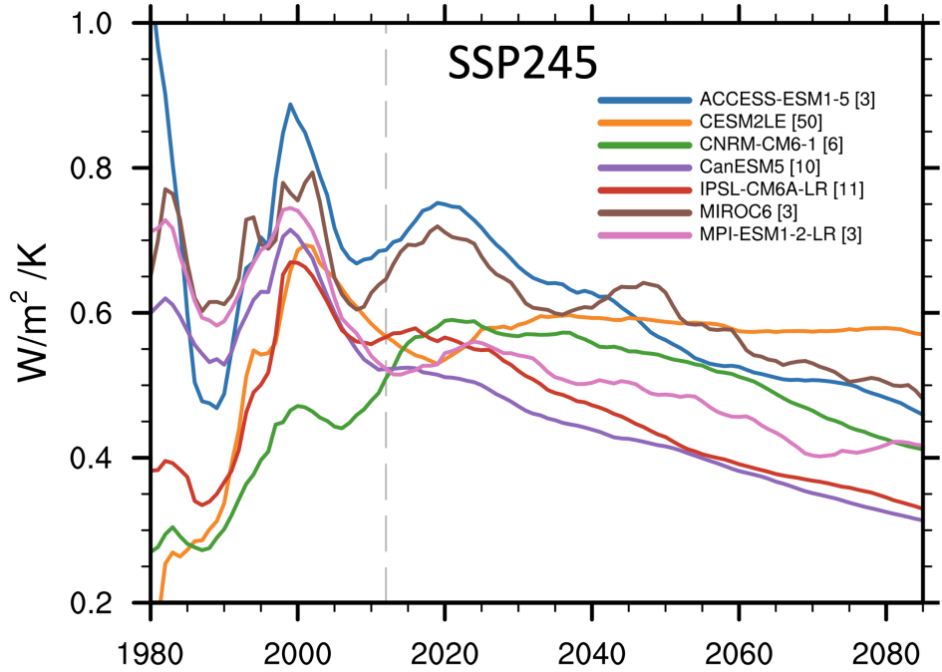

**Figure S7.** Time evolution of OHUE from historical runs and Shared Socioeconomic Pathways 2-4.5 (SSP245). The results represent ensemble mean from individual models. Similar to Figure 1a, the x-axis shows the centered year of the 20-year sliding window. The ensemble size used is determined by the ensemble size of the corresponding SSP245 runs, which are shown in the square brackets.